\documentclass[aps,twocolumn,epsfig,prb]{revtex4}

\newcommand{\bsigma}{\mbox{\boldmath $\sigma$}}

\usepackage{amsmath}
\usepackage{graphicx}

\begin{document}

\title{Aharanov-Bohm effect for the edge states of zigzag carbon
nanotubes}

\author{K. Sasaki}
\email[Email address: ]{sasaken@flex.phys.tohoku.ac.jp}
\affiliation{Department of Physics, Tohoku University and CREST, JST,
Sendai 980-8578, Japan}

\author{M. Suzuki}
\affiliation{Department of Physics, Tohoku University and CREST, JST,
Sendai 980-8578, Japan}

\author{R. Saito}
\affiliation{Department of Physics, Tohoku University and CREST, JST,
Sendai 980-8578, Japan}

\date{\today}
 
\begin{abstract}
 Two delocalized states of metallic zigzag carbon nanotubes
 near the Dirac point 
 can be localized by the Aharanov-Bohm magnetic field around 20 Tesla.
 The dependence of the localization 
 on the length and diameter of the nanotubes 
 shows that the localization-delocalization transition can be observed
 for 2 nm diameter tube.
 The mechanism of the localization is explained 
 in terms of the deformation-induced gauge field,
 which shows a topological nature of the localization.
 The transition from the delocalized states to the localized states
 can be observed by scanning tunneling microscopy and spectroscopy. 
 A similarity between the transition and the spin Hall effect
 is discussed.
\end{abstract}

\maketitle

\section{introduction}\label{sec:intro}

The electronic properties of graphene
have attracted much attention from various points of view.
It is found that 
graphene shows 
the integer quantum Hall effect~\cite{novoselov05,zhang05} 
and dissipationless supercurrent.~\cite{heersche07}
These effects are attributed to
the energy band structure of graphene which
consists of two Dirac cones at the K and K' points 
in the $k$-space.
The dynamics of electrons around each Dirac point 
is approximated by the Weyl equation, 
which describes a ``massless'' particle.
The ``massless'' particle never stop 
and the wave function is generally extended.
However, electrons can be localized near the zigzag edge of graphene,
which are called the edge states.~\cite{fujita96}
The appearance of the edge states is sensitive to the shape of the edge,
that is, the zigzag edge induces the edge states
while the armchair edge does not.
Since the energy dispersion of the edge states
as a function of the wave vector along the edge direction
appears near the Dirac points, 
the local electronic properties such as ferromagnetism~\cite{fujita96} 
and superconductivity~\cite{sasaki06super}
near the zigzag edge are proposed in terms of the edge states.
The edge states exist near the zigzag end 
of a single-wall carbon nanotube, too, because
a carbon nanotube is a graphene sheet wrapped into a cylinder.

The energy dispersion relation of the zigzag edge states 
appears only between the two Dirac points,
and the localization length ($\xi$) 
of the edge state depends on the distance
from the Dirac point in the $k$-space.
In particular, at the center of the two Dirac points, 
the wave function of the edge state has amplitude only 
at the edge sites ($\xi=0$).
While $\xi$ becomes infinite at the Dirac points where
the edge states connect to extended states continuously. 
Thus, by changing $k$ due to the Aharanov-Bohm (AB) effect
for the magnetic flux penetrating a hollow core of
nanotube,~\cite{ajiki93}  
an extended state at the Dirac point can be transfered into 
an edge state and vice versa 
(localization-delocalization (LD) transition).
In the previous paper,~\cite{sasaki05prb}
we showed that the LD transition is possible for large diameter
zigzag nanotubes and 
it can be observed by the conductance measurement.
In this paper,
we first show analytical calculations of the length and diameter 
dependence of the LD transition,
and then try to explain the phenomena intuitively 
using a continuous model.
We will show that the LD transition can be observed by 
scanning tunneling microscopy (STM) 
and spectroscopy (STS) measurements in the presence of magnetic field
around 20 Tesla.

Since the edge states exist near the Fermi energy, 
the real-space image of the edge states is observed by STM
experiments.~\cite{klusek00,giunta01,niimi05,kobayashi05,niimi06,kobayashi06}  
The local density of state (LDOS) is observed by STS 
at a step edge of the zigzag type on a vicinal surface of
graphite.~\cite{klusek00,niimi05,kobayashi05,niimi06,kobayashi06}
The cylindrical structure of carbon nanotubes 
is suitable for the study of the AB effect. 
The AB oscillations and 
the period of the fundamental unit of magnetic flux ($\Phi_0$)
were observed in multi-wall nanotubes.~\cite{bachtold99,coskun04} 
Since AB flux breaks time-reversal symmetry, 
a splitting of the degenerated van Hove singularity 
for K and K' points is observed.~\cite{ajiki93,roche00,b782} 
The splitting was observed 
as a shift of the first-subband magneto-absorption peak 
in semiconducting single-wall nanotubes~\cite{zaric04}
and as a splitting of the peak position of the van Hove singularities 
in the conductance measurement.~\cite{minot04} 
These experiments are intended to observe the AB effect
for the extended states near the Fermi level.
The AB measurement by STM/STS for the edge states 
not only gives a direct evidence of the edge states 
in zigzag carbon nanotubes
but also can clarify the topological property of the edge states.

An important property of the edge state is that
the wave function of the edge state has
an amplitude only on one of the two sublattices (A
and B) in the hexagonal lattice. When we consider
a pseudo 1/2 spin whose up and down spins represent the relative
amplitude on the A and B sublattices, respectively, 
an edge state can be described by a pseudo-spin polarized state 
accumulated at the edge. 
This situation is similar to the spin Hall
effect~\cite{ShuichiMurakami09052003,sinova:126603,PhysRevLett.83.1834} 
in which the spin-orbit interaction induces the spin polarization
at the edge of semiconductor materials by 
``the Lorentz force for spin'' in the presence of the electronic
current. 
In this paper, 
we will show that a similar Lorentz force acts
for the pseudo spin in the graphene system 
in which the lattice defects can be understood 
by the time-reversal-symmetric gauge field and 
by corresponding pseudo-magnetic field. 

In Sec.~\ref{sec:AB}, we show the AB effect for the edge states. 
In Sec.~\ref{sec:cmodel}, 
the pseudo spin and corresponding Hamiltonian are defined and 
the Lorentz force for the pseudo spin is derived.
In Sec.~\ref{sec:discussion}, discussion and summary will be given.

\section{AB effect for the edge states}\label{sec:AB}

Here, we define the wave number around and along the axis of a tube 
as $k_{\rm c}$ and $k_{\rm t}$, respectively.
Because of the periodic boundary condition 
around the axis for a $(n,0)$ zigzag nanotube, 
$k_{\rm c}$ is discrete as $k_{\rm c} = 2\pi p/|{\bf C}_h|$ 
($p$ is integer) where $|{\bf C}_h|=na$ and
$a=0.246$ nm is the lattice constant.
$k_{\rm t}$ is also quantized by the boundary condition 
in the direction of the axis of the tube for a finite length $L$.
In the previous paper,~\cite{sasaki05prb}
we give the boundary condition for $k_{\rm t}$ as follows,
\begin{align}
 -2 \left( 1 + \frac{\alpha}{n^2} \right) 
 \cos \left( \frac{k_{\rm c} a}{2} \right) = 
 \frac{\sin \left( k_{\rm t} (L+\ell) \right)}{\sin \left( k_{\rm t} (L+2\ell)\right)},
 \label{eq:bc}
\end{align}
where $2\ell\equiv \sqrt{3}a$ is the unit length 
in the direction of the axis
and $\alpha$ is a parameter representing the curvature effect.
The energy for $pi$-band is given by
\begin{align}
 & E(k_{\rm c},k_{\rm t}) = \pm \gamma_0 \sqrt{g(k_{\rm c})^2 + 2g(k_{\rm c})\cos(k_{\rm t} \ell) +1}, 
 \label{eq:energy}\\
 & {\rm where} \ \ g(k_{\rm c})\equiv 2 \left( 1 + \frac{\alpha}{n^2} \right)  
 \cos \left( \frac{k_{\rm c} a}{2} \right)
 \nonumber
\end{align}
and $\gamma_0$ ($\approx 3$ eV) 
is the nearest neighbor hopping integral.

First we consider the case of $\alpha=0$ in Eqs.~(\ref{eq:bc}) 
and (\ref{eq:energy}) for simplicity.
Then we will discuss the case for $\alpha=\pi^2/8$ 
which is derived previously.~\cite{sasaki05prb} 
For the K point $(k_{\rm c},k_{\rm t})=(4\pi/3a,0)$
(K' point $(k_{\rm c},k_{\rm t})=(2\pi/3a,\pi/\ell)$), 
we get $g(k_{\rm c})=-1$ ($g(k_{\rm c})=1$) and 
$E(k_{\rm c},k_{\rm t})=0$ (Dirac points).

\begin{figure}[htbp]
 \begin{center}
  \includegraphics[scale=0.4]{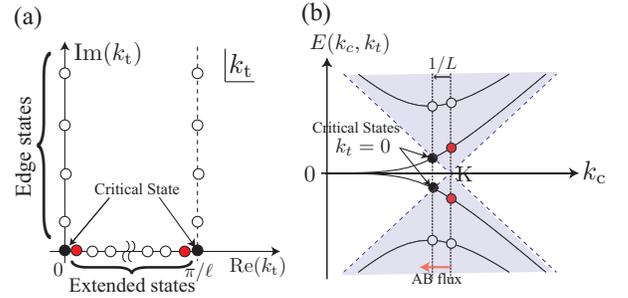}
 \end{center}
 \caption{(color online) 
 (a) In the complex plane of $k_{\rm t}$, 
 the extended states have real $k_{\rm t}$, while
 the edge states have imaginary part in $k_{\rm t}$.
 $k_{\rm t}=0$ (and $k_{\rm t}=\pi/\ell$) is the critical state.
 (b) The energy band structure $E(k_{\rm c},k_{\rm t})$ near the K-point
 is plotted as a function of $k_{\rm c}$ where
 the $k_{\rm t}$ axis is perpendicular to the plane.
 Each dispersion curve corresponds to a different value of $k_{\rm t}$.
 The two states represented by the red circles 
 can go out of the surface of the Dirac cones at the critical states
 (represented by solid black circles)
 by means of the AB flux
 and go into the localized region.
 }
 \label{fig:critical}
\end{figure}

Depending on the value of $k_{\rm c}$,
Eq.~(\ref{eq:bc}) has real and imaginary solutions for $k_{\rm t}$ 
corresponding to the extended and the edge states, respectively. 
It can be shown that 
the edge states appear when $2\pi/3a<k_{\rm c}<4\pi/3a$
($|g(k_{\rm c})| < 1$),
and that $k_{\rm t}$ for the edge state satisfies
\begin{align}
 k_{\rm t} = 
 \begin{cases}
  \displaystyle
  \frac{\pi}{\ell} + \frac{i}{\xi(k_{\rm c})} &  \displaystyle
  \left( \frac{2\pi}{3a} < k_{\rm c} < \frac{\pi}{a} \right) \\
  \displaystyle \frac{i}{\xi(k_{\rm c})} &  \displaystyle
  \left( \frac{\pi}{a} <k_{\rm c}< \frac{4\pi}{3a} \right)
 \end{cases}
\end{align}
where $\xi(k_{\rm c})$ denotes the localization length
of the edge state defined by
$\xi(k_{\rm c})=-\ell/\ln(|g(k_{\rm c})|)$.~\cite{sasaki05prb}
At $k_{\rm t} = 0$ or $\pi/\ell$ (or when $\xi(k_{\rm c})$ becomes $\infty$), 
we have a discontinuous change of $k_{\rm t}$
(see Fig.~\ref{fig:critical}(a)).
The states for $k_{\rm t}=0$ or $k_{\rm t}=\pi/\ell$ can be called 
``critical states'' since they can be regarded both
as an extended state ($k_{\rm t}$ is a real number)
and as a localized state 
with infinite localization length ($\xi\to \infty$).
By substituting $k_{\rm t}=0$ into Eq.~(\ref{eq:bc}),
we obtain
\begin{align}
 k_{\rm c}^{\rm critical}
 &= \frac{2}{a} 
 \arccos \left( -\frac{1}{2} \frac{L+\ell}{L+2\ell} \right) \nonumber \\
 &\approx 
 \frac{4\pi}{3a} - \frac{1}{L}, \ \ (L \gg \ell).
 \label{eq:kc_cri}
\end{align}
$k^{\rm critical}_{\rm c}$ corresponds to the K point 
($g(k^{\rm critical}_{\rm c})=-1$) 
when $L\to \infty$.
Similarly, $k_{\rm t}=\pi/\ell$ 
gives $k^{\rm critical}_{\rm c} \approx 2\pi/3a+1/L$ ($L\gg \ell$)
and $k_{\rm c}$ becomes the K' point in the limit of $L\to \infty$.
In Fig.~\ref{fig:critical}(b), 
we plot Eq.~(\ref{eq:energy}) around the K point
as a function of $k_{\rm c}$ 
where $k_{\rm t}$ is determined by Eq.~(\ref{eq:bc}).
$k_{\rm t}$ is a real number in the shaded region 
while $k_{\rm t}$ is a complex number 
outside of the shaded region (localized region).
The critical states are denoted by the solid black circles.
For the critical states, 
we obtain $g(k^{\rm critical}_{\rm c})\approx -1+\ell/L$.
By putting this into Eq.~(\ref{eq:energy}), 
we obtain the energy eigenvalues of the critical states
as $E(k^{\rm critical}_{\rm c},0)=\pm \gamma_0 \ell/L$.
The critical states are located on the inter section made by 
the surface of the Dirac cone and the plane of $k_{\rm t}=0$.
The inter section is denoted by the dashed lines 
in Fig.~\ref{fig:critical}(b).


In case of the metallic zigzag nanotubes
($n=3q$ where $q$ is integer), 
one of the discrete value of $k_{\rm c}(=2\pi p/na)$ 
intersects the K point at $k_{\rm c}=4\pi/3a$.
The AB flux along the axis of a tube
shifts the electronic state from the K point to 
\begin{align}
 k_{\rm c}(n_\Phi) = \frac{4\pi}{3a} - \frac{2\pi n_\Phi}{|{\bf C}_h|},
 \label{eq:kc_gauge}
\end{align}
where $n_\Phi$ is number of the flux quantum.
In the presence of a uniform magnetic field of $B$[T], 
$n_\Phi$ for (n,0) zigzag tube is expressed by
\begin{align}
 n_\Phi \equiv \frac{BS}{\Phi_0}= \frac{B}{B_1}n^2,
 \label{eq:flux}
\end{align}
where $S=\pi(na/2\pi)^2$ is the cross sectional area of the nanotube,
$\Phi_0=4.1\times 10^{5}$[T \AA$^2$], and 
$B_1=8.5\times 10^5$[T].
Thus, $n_\Phi=1$ (or $\Phi_0$) corresponds to 
$B\approx 1000$[T] for $n=30$ 
(diameter of the tube: $d_t=|{\bf C}_h|/\pi$ is 2.35 nm).
1000[T] is beyond an accessible magnetic field.
However, 
the transition from an extended state to an edge state does not 
require such a strong magnetic filed even for $d_t \approx 2$ nm.
Comparing Eq.~(\ref{eq:kc_cri}) with Eq.~(\ref{eq:kc_gauge}),
we see that the magnetic field which shifts from the K point 
to the critical state is proportional to $|{\bf C}_h|/L$ as
\begin{align}
 n_\Phi^{\rm critical}=\frac{|{\bf C}_h|}{2\pi L},\ \ \ 
 \left({\rm or} \ \ 
 L^{\rm critical}= \frac{B_1}{B} \frac{a^2}{2\pi^2 d_t} \right).
 \label{eq:a}
\end{align}
Since $L \gg |{\bf C}_h|$ holds for nanotubes,
the magnetic field for the critical state 
becomes much smaller than 1000[T].
For example, 
corresponding magnetic field becomes 10[T] 
when $L$ is larger than $L^{\rm critical}=230$ nm 
for a $(15,0)$ zigzag nanotube.

Although the critical states exist at the K' point,
the critical states at the K and K' points 
do not occur simultaneously. 
It is because that 
the critical states at the K' point appears
for $1-n_\Phi$ flux.

\begin{figure}[htbp]
 \begin{center}
  \includegraphics[scale=0.5]{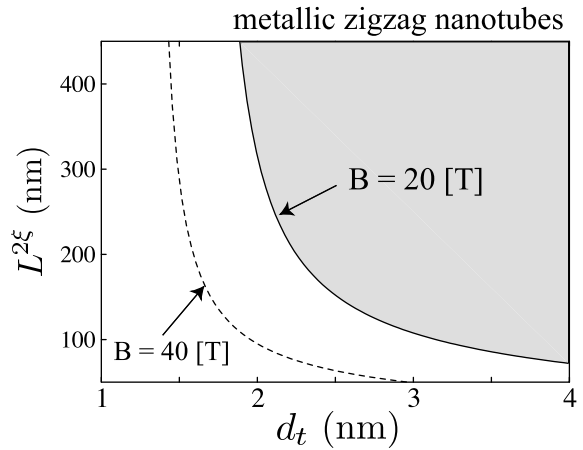}
 \end{center}
 \caption{
 The minimum length ($L^{2\xi}$) and diameter ($d_t$) for obtaining
 the critical states by AB effect for 20[T] (solid curve)
 and 40[T] (dashed curve).
 The curves diverge at
 $d_t=4.2(\alpha/B[{\rm T}])^{1/3}$ nm.
 }
 \label{fig:critical_region}
\end{figure}

When the curvature effect ($\alpha\ne 0$) is included, 
the expression for $k_{\rm c}$ (Eq.~(\ref{eq:kc_cri}))
is modified.
From Eq.~(\ref{eq:bc}), we obtain
\begin{align}
 k_{\rm c}^{\rm critical} \approx \frac{4\pi}{3a} - \frac{1}{L}
 -\frac{\alpha}{n^2 \ell}, \ \ (L \gg \ell).
 \label{eq:kc}
\end{align}
By comparing Eq.~(\ref{eq:kc_cri}) with Eq.~(\ref{eq:kc}), 
we see that the curvature effect increases the distance 
between the electronic state and the critical state
by $\alpha/n^2 \ell$. 
Then, comparing Eq.~(\ref{eq:kc}) with Eq.~(\ref{eq:kc_gauge}),
we see that the magnetic field which shifts from the K point 
to the critical state becomes
\begin{align}
 & n_\Phi^{\rm critical} = \frac{|{\bf C}_h|}{2\pi L} 
 + \frac{\alpha}{2\pi n}\frac{a}{\ell}, \nonumber \\ 
 & \left( {\rm or} \ \
 L^{\rm critical} =
 \frac{1}{\displaystyle \frac{B}{B_1}\frac{2\pi^2 d_t}{a^2}
 - \frac{\alpha}{\pi^2} \frac{a^2}{d_t^2 \ell}}\right).
 \label{eq:cri_met}
\end{align}
Since the $L^{\rm critical}$ corresponding to $n_\Phi^{\rm critical}$
becomes infinite when
$d_t=4.2(\alpha/B[{\rm T}])^{1/3}$nm in Eq.~(\ref{eq:cri_met}),
$d_t$ must be larger than this value 
to reach the critical states for a finite length nanotube.
For example, $d_t$ must be larger than 1.66 nm for $B=20$[T].
It is important to note that 
we do not need to discuss the case that the localization length $\xi$ is
larger than $L$.
In order to observe the critical transition in experiments,
it is sufficient to get $\xi=L/2$.
By putting $k_{\rm t} = i/\xi$ with $\xi=L/2$ to Eq.~(\ref{eq:bc}),
we obtain 
\begin{align}
 L^{2\xi} =
 \frac{2\coth(2)}{\displaystyle \frac{B}{B_1}\frac{2\pi^2 d_t}{a^2}
 - \frac{\alpha}{\pi^2} \frac{a^2}{d_t^2 \ell}},
 \label{eq:L(d)}
\end{align}
in stead of Eq.~(\ref{eq:cri_met}).
The finite localization length appears as a factor
of $2\coth(2)\approx 2.1$.
In Fig.~\ref{fig:critical_region}, 
we plot $L^{2\xi}$ in Eq.~(\ref{eq:L(d)}) as a function of $d$
for $B=20$[T] and 40[T] for metallic zigzag nanotubes ($n=3q$).
The shaded area in Fig.~\ref{fig:critical_region}
corresponds to possible length and diameter 
to observe the LD transition at $B=20$[T] or lower.

In the case of semiconducting nanotubes, 
$k_{\rm c}$ does not exist at the K point, 
which requires a large $B$ as is shown below.
Semiconducting zigzag nanotubes are divided into 
type I ($2n=3p+1$) and type II ($2n=3p-1$) where $p$ is
integer.~\cite{saito:153413} 
Since $p=2n/3 -1/3$ holds for type I and
$p=2n/3 +1/3$ for type II,
we have the electronic states at 
\begin{align}
 k_{\rm c} = \frac{2\pi p}{na} = \frac{4\pi}{3a}\mp\frac{2\pi}{3na},
\end{align}
where minus (plus) sign in front of $2\pi/3na$ is for type I (II).
Then, in the presence of the magnetic field, we have
\begin{align}
 k_{\rm c} = \frac{4\pi}{3a}\mp\frac{2\pi}{3na}
 -\frac{2\pi n_\Phi}{|{\bf C}_h|}.
\end{align}
The electronic states of type I 
which are located closest to the critical states 
are the edge states when $B=0$,
and become the extended states by applying a magnetic field
(delocalization).
Those for type II are the extended states when $B=0$,
and become the edge states by $B$.
Comparing this with Eq.~(\ref{eq:kc}), we see that
\begin{align}
 n_\Phi^{\rm critical} = \frac{|{\bf C}_h|}{2\pi L} 
 + \frac{\alpha}{2\pi n}\frac{a}{\ell} \mp \frac{1}{3}
\end{align}
is necessary to obtain the critical states 
in the semiconducting nanotubes.
Due to the last term ($\mp 1/3$),
we need a large diameter tube of order of 10 nm 
in order to see the critical states by an accessible magnetic field.
In this respect, semiconducting tubes 
are not suitable to observe the critical states.

\section{Continuous model}\label{sec:cmodel}

In the previous section, 
we have shown within the tight-binding model
that the extended states are changed into the edge states
through the critical states by the AB flux.
The existence of the edge states and 
the critical states at $k_{\rm t}=0$ (or $k_{\rm t}=\pi/\ell$)
is originated from the boundary condition of Eq.~(\ref{eq:bc}).
In this section, 
we try to explain the LD transition using a continuous model, 
which is useful to understand the phenomena intuitively.

\begin{figure}[htbp]
 \begin{center}
  \includegraphics[scale=0.4]{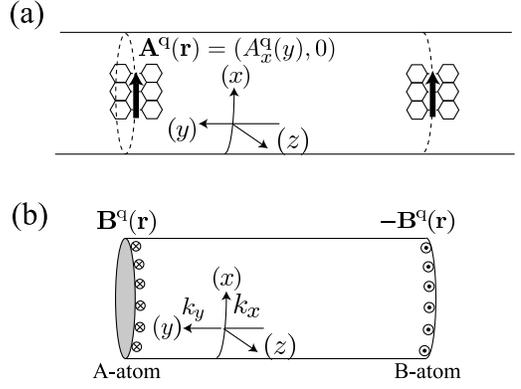}
 \end{center}
 \caption{
 (a) A zigzag nanotube is obtained from a periodic tube 
 by cutting the bonds at the zigzag edge.
 The change of the hopping integral due to the bond-cutting
 appears as the deformation-induced gauge field,
 ${\bf A}^{\rm q}({\bf r})$, in the Weyl equation. 
 (b) The deformation-induced ``magnetic'' field 
 ${\bf B}^{\rm q}({\bf r})$ appears at the zigzag edge.
 For a zigzag edge consisting of A-atoms, ${\bf B}^{\rm q}({\bf r})$
 points to the negative $z$-direction, while 
 it is positive direction for a zigzag edge consisting of B-atoms.
 ${\bf B}^{\rm q}({\bf r})$ does not depends on $x$ due to the
 translational symmetry.
 }
 \label{fig:gauge}
\end{figure}

In the continuous model for nanotubes,
the modification of hopping integral 
due to a local lattice deformation 
appears as a deformation-induced gauge field, 
${\bf A}^{\rm q}({\bf r})$, 
in the Weyl equation,~\cite{kane97,sasaki05}
${\cal H}_{\rm K} \psi_{\rm K}({\bf r}) = E \psi_{\rm K}({\bf r})$ 
where
\begin{align}
 {\cal H}_{\rm K} = 
 v_{\rm F} \bsigma \cdot ({\bf p}+{\bf A}^{\rm q}({\bf r})),
 \label{eq:weyl}
\end{align}
$v_F$ is the Fermi velocity, and
$\bsigma=(\sigma_x,\sigma_y)$ is the Pauli spin matrix.
The wave function 
$\psi_{\rm K}({\bf r})={}^t(\psi_{\rm A}({\bf r}),\psi_{\rm B}({\bf r}))$ 
has two components
which represent the wave functions for two atoms (A and B) 
in the unit cell. 
As we mention in Introduction,
since the two component wave function 
is similar to the electron spin,
we call $\psi_{\rm K}({\bf r})$ the pseudo-spin.
${\bf A}^{\rm q}({\bf r})$ is different from 
the electro-magnetic gauge field ${\bf A}^{\rm em}({\bf r})$
in the sense that the ${\bf A}^{\rm q}({\bf r})$ holds time-reversal
symmetry.~\cite{sasaki05}
By considering a bond-cutting procedure at the edge
as an extreme case of the deformation (see Fig.~\ref{fig:gauge}(a)),
we showed that
the deformation-induced ``magnetic'' field,
${\bf B}^{\rm q}({\bf r})\equiv \nabla \times {\bf A}^{\rm q}({\bf r})$, 
appears at the zigzag edge (Fig.~\ref{fig:gauge}(b)).
The ${\bf B}^{\rm q}({\bf r})$ field
represents the boundary condition for the zigzag edge
(Eq.~(\ref{eq:bc}))
and explains the occurrence of the edge states.~\cite{sasaki06jpsj}
Since ${\bf A}^{\rm q}({\bf r})$ is a vector 
which lies on the surface of the graphene
and has only $x$-component,~\cite{sasaki06jpsj}
${\bf B}^{\rm q}({\bf r}) = (0,0,B_z^{\rm q}({\bf r}))$ 
is normal to the nanotube surface ($z$-direction).
The direction of ${\bf B}^{\rm q}({\bf r})$ field
becomes opposite for the both ends of a zigzag nanotube.
That is, for a zigzag edge consisting of A-atoms,
we have ${\bf B}^{\rm q}({\bf r})$, 
while for another zigzag edge consisting of B-atoms,
we have $-{\bf B}^{\rm q}({\bf r})$ (see Fig.~\ref{fig:gauge}).
Since Eq.~(\ref{eq:weyl}) does not depend on $t$ explicitly,
the energy of the system is conserved.

\begin{figure}[htbp]
 \begin{center}
  \includegraphics[scale=0.4]{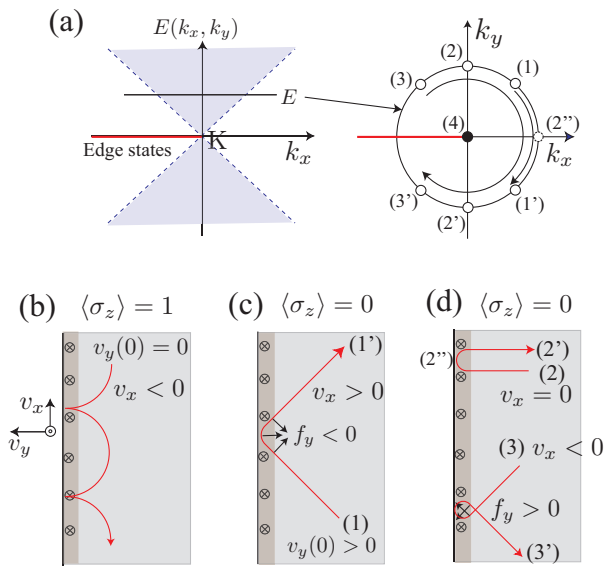}
 \end{center}
 \caption{(color online) (a) 
 We consider the scattering processes for the initial states
 at (1)$\sim$(3) in the $k$-space with energy $E\ne 0$,
 and the initial state at (4) with $E=0$.
 (b) The motion of the edge state is the cyclotron motion.
 (c,d) The velocity ${\bf v}=(v_x,v_y)$ causes the Lorentz force,
 $-{\bf v}\times {\bf B}^{\rm q}({\bf r})$, at the edge sites.
 The force can attract or reflect an incident electron depending on 
 the sign of $v_{x}$.
 } 
 \label{fig:scatter}
\end{figure}

We consider the particle velocity, ${\bf v} (=(v_x,v_y))$, 
defined by
\begin{align}
 {\bf v} \equiv \frac{d{\bf r}}{dt} = \frac{1}{i\hbar}
 \left[{\bf r}, {\cal H}_{\rm K} \right].
 \label{eq:eom_r}
\end{align}
Using Eq.~(\ref{eq:weyl}), 
we get ${\bf v}=v_{\rm F}\bsigma$.
For a Dirac particle with momentum
${\bf p}(=|{\bf p}|{\hat {\bf p}})$, 
we obtain $\langle {\bf v} \rangle=v_{\rm F}\hat{{\bf p}}$.
The motion of the edge states can be understood 
from the time-derivative of $v_x$ and $v_y$:
\begin{align}
 \begin{split}
  & \frac{dv_x}{dt} = \frac{1}{i\hbar}
 [v_F\sigma_x,{\cal H}_{\rm K}]=
 \frac{2v_F^2}{\hbar} \sigma_z \pi_y, \\
 & \frac{dv_y}{dt} = \frac{1}{i\hbar}
 [v_F\sigma_y,{\cal H}_{\rm K}]=
 -\frac{2v_F^2}{\hbar} \sigma_z \pi_x,
 \end{split}
 \label{eq:eom_edge}
\end{align}
where ${\boldsymbol{\pi}}=(\pi_x,\pi_y)(
\equiv {\bf p}+{\bf A}^{\rm q}({\bf r}))$ 
is the kinematical momentum.
The wave function of the edge state ($\psi^{\rm edge}({\bf r})$)
is polarized in terms of the pseudo-spin. 
In fact,
$\psi^{\rm edge}_{\rm K}({\bf r}) \propto {}^t(1,0)$ 
near the zigzag edge consisting of A-atoms
and $\psi^{\rm edge}_{\rm K}({\bf r}) \propto {}^t(0,1)$ 
near the zigzag edge consisting of B-atoms.
We have $\langle \sigma_z \rangle =\pm 1$
for pseudo-spin polarized states.
Then, by putting 
${\bf r}=r^{\rm cyc}(\sin \omega t,\cos \omega t)$
and ${\bf p}=|{\bf p}|(\cos \omega t,-\sin \omega t)$
into $\langle d{\bf r}/dt \rangle = v_{\rm F} \hat{\bf p}$ and
Eq.~(\ref{eq:eom_edge}), 
we get the cyclotron motion with the cyclotron radius,
$r^{\rm cyc} = 2\hbar/|{\bf p}|$ in the absence of 
${\bf A}^{\rm q}({\bf r})$ field.
Since only the $x$-component of the ${\bf A}^{\rm q}({\bf r})$ field 
appears at the zigzag boundary ($y=0$),~\cite{sasaki06jpsj} 
we have $\pi_x= p_x + A_x^{\rm q}(y)$ and $\pi_y = p_y$.
Thus, for the initial pseudo-spin polarized state with $p_y=0$, 
the state follows the cyclotron motion and 
$v_y$ changes the sign at the boundary due to $A_x^{\rm q}(y)$
as shown in Fig.~\ref{fig:scatter}(b).
The corresponding states are the edge states.
In fact, the localization length of the edge states
is calculated as $\xi = \hbar/|p_x|$ in the continuous
model,~\cite{sasaki06jpsj} 
which is the same as the $r^{\rm cyc}/2$ for $p_y=0$.
The cyclotron motion of the edge state 
is the eigenstate of the total angular momentum 
($J_z=\hat{\bf z}\cdot ({\bf r}\times {\boldsymbol{\pi}}) +
(\hbar/2)\sigma_z$) 
with eigenvalue $J_z=\mp 3\hbar/2$.

To see the correspondence more in detail, we need to consider 
how the pseudo-spin polarization is achieved by the 
${\bf A}^{\rm q}({\bf r})$ field.
Time-evolution of the $\sigma_z$ is given by
\begin{align}
 \frac{d\sigma_z}{dt} 
 = \frac{1}{i\hbar}[\sigma_z, {\cal H}_{\rm K}]
 = \frac{2v_F}{\hbar} \bsigma \cdot 
 (\hat{\bf z}\times {\boldsymbol{\pi}}).
\end{align}
In the absence of ${\bf A}^{\rm q}({\bf r})$, 
since $\bsigma$ and ${\bf p}$ are parallel or anti-parallel
for the extended states (helicity),~\cite{ando05}
we have $\langle \sigma_z \rangle=0$ and 
$\langle d\sigma_z/dt \rangle =0$, and 
Eq.~(\ref{eq:eom_edge}) does not give the cyclotron motion.
On the other hand, at the zigzag edge, 
$d\sigma_z/dt \ne 0$ since ${\bf A}^{\rm q}({\bf r}) \ne 0$.
Moreover, it can be shown that
$d^2 \sigma_z/dt^2 = -4 \sigma_z({\cal H}_{\rm K}/\hbar)^2
+ (2v_{\rm F}^2/\hbar) \hat{{\bf z}}\cdot {\bf B}^{\rm q}({\bf r})$.
Thus, the pseudo-spin (or the edge states)
is accumulated at the zigzag edge.

The scattering process at the zigzag edge 
for the extended (pseudo-spin unpolarized, $\langle \sigma_z \rangle=0$)
states can be understood by 
the equation of motion of ${\boldsymbol{\pi}}$, which is given by 
\begin{align}
 \frac{d{\boldsymbol{\pi}}}{dt} = \frac{1}{i\hbar}
 \left[{\boldsymbol{\pi}}, {\cal H}_{\rm K} \right]
 = - v_F \bsigma \times {\bf B}^{\rm q}({\bf r}).
 \label{eq:def_F}
\end{align}
The right-hand side of Eq.~(\ref{eq:def_F}) shows 
that the Dirac particle undergoes a ``Lorentz force'': 
\begin{align}
 {\bf f}({\bf r})=
 -{\bf v} \times {\bf B}^{\rm q}({\bf r}).
 \label{eq:force}
\end{align}
The Lorentz force rotates the momentum ${\bf p}$ 
of the incident Dirac particle at the zigzag boundary.
Due to the helicity conservation, the pseudo-spin and ${\bf p}$ 
are parallel ($\langle {\bf v} \rangle =v_{\rm F} \hat{\bf p}$) 
in the scattering process.
We consider the time-evolution of the following four initial states
specified by ${\bf v}=(v_x,v_y)$ as
(1) $(v_x>0,v_y>0)$, 
(2) $(0,v_y>0)$, 
(3) $(v_x<0,v_y>0)$, and
(4) $(0,0)$.
The corresponding position of each initial state in the $k$-space is 
shown in Fig.~\ref{fig:scatter}(a).
Hereafter, we denote the $(x,y)$ components of ${\bf f}$ as
$(f_x,f_y)$. 

First, we consider the scattering process for (1).
When $v_x > 0$, we have $f_y<0$ from Eq.~(\ref{eq:force})
and ${\bf B}^{\rm q}({\bf r})$ field reflects the electron 
at the zigzag edge.
The trajectory of the Dirac particle is 
shown in Fig.~\ref{fig:scatter}(c),
and the final state is given by (1').
Due to the energy conservation, the time-evolution of ${\bf p}$ 
is restricted on the circle with radius $|{\bf p}|$
in the $k$-space.

Next, we consider the initial state of (2)
(see Fig.~\ref{fig:scatter}(d)).
In this case, the state is reflected by the zigzag edge
and changes the sign of $v_y$, and
the final state is given by (2').
The time-evolution of the ${\bf v}$ 
in this scattering process is as follows.
First, 
the ${\bf B}^{\rm q}({\bf r})$ field changes $(0,v_y)$ to
$(f_x dt,0)$ ((2'') in Fig.~\ref{fig:scatter}(a)) 
in a very short period ($dt$).
Then, the velocity of the virtual state 
is rotated by ${\bf B}^{\rm q}({\bf r})$ field again,
and the final state becomes $(0,-v_y)$.
This explains that the state moves in the clockwise direction in
the $k$-space (see Fig.~\ref{fig:scatter}(a)) and 
reaches the final state.
The presence of ${\bf B}^{\rm q}({\bf r})$ field 
gives rise to a phase shift in the scattering process, 
and yields the backward scattering.
According to the absence of the backward scattering mechanism,~\cite{ando98}
the Berry's phase shift of $\pi$ between the two scattered waves
corresponding to the clockwise and anticlockwise rotation in the $k$-space,
cancels the back-scattering amplitude.
The ${\bf B}^{\rm q}({\bf r})$ field selects only the clockwise motion
in the $k$-space and recovers the backward scattering
at the zigzag edge.

For the initial state of (3), 
the direction of $f_y$ becomes $f_y >0$ 
and then the ${\bf B}^{\rm q}({\bf r})$ field
tends to trap the electrons.
However, due to the energy conservation, 
the electron can escape from the edge and 
the final state is given by (3').
The trajectory of the Dirac particle is 
shown in Fig.~\ref{fig:scatter}(d).

Finally,
for the initial state of (4) (i.e., particle at the Dirac point), 
the particle is not affected by ${\bf B}^{\rm q}({\bf r})$ field
(${\bf f}={\bf 0}$).
The AB flux along the axis of a tube gives a finite $v_x$
and the ${\bf B}^{\rm q}({\bf r})$ field produces the 
non-vanishing Lorentz force.
Then, the ${\bf B}^{\rm q}({\bf r})$ field attracts
the state with $v_y=0$ at the zigzag edge if $v_x <0$.
The state at the Dirac point is unstable against the AB flux
and undergoes the LD transition.
This state is nothing but the critical state
that we discussed in this paper.

\section{discussion}\label{sec:discussion}


It is interesting that 
the localization phenomena discussed in this paper 
is analogous to the spin Hall effect
(SHE).~\cite{ShuichiMurakami09052003,sinova:126603,PhysRevLett.83.1834}
In the SHE, the spin current is accumulated near the edges
of semiconductor materials by the electric field applied along the edge.
Since the time derivative of the AB flux 
gives an electronic field along the zigzag edge,
the physical situation discussed in this paper is similar to 
that of the SHE. 
The wave function of the edge state in graphene
is polarized in terms of the pseudo-spin. 
Since the extended state is a pseudo-spin unpolarized state,
the pseudo-spin is accumulated by the localization.
Thus, by neglecting the difference between 
the (real) spin in the SHE and 
the pseudo-spin, 
the situations of these systems are quite similar to each other.

\begin{table*}[htbp]
 \caption{\label{tab:analogy}Analogy between graphene and SHE} 
 \begin{ruledtabular}
  \begin{tabular}{lll}
   & {\bf Graphene} & {\bf SHE} \\
   Wave function & Pseudo-spin & Spin \\
   Hamiltonian & 
   ${\cal H}_{\rm K}= v_{\rm F} \bsigma \cdot {\boldsymbol{\pi}}$ &
   ${\cal H}_{\rm so}=-(\lambda/\hbar) \bsigma \cdot (\hat{\bf z} \times
   {\bf p})$ \\
   $d\sigma_z/dt$ & 
   $(2v_{\rm F}/\hbar) \bsigma \cdot (\hat{\bf z} \times
   {\boldsymbol{\pi}})$ &  
   $(2\lambda/\hbar^2) \ \bsigma \cdot {\bf p}$ \\
   $d^2\sigma_z/dt^2$ & 
   $-4 \sigma_z({\cal H}_{\rm K}/\hbar)^2
   + (2v_{\rm F}^2/\hbar) B_z^{\rm q}({\bf r})$ &
   $-4 \sigma_z({\cal H}_{\rm so}/\hbar)^2$
   \\
  \end{tabular}
 \end{ruledtabular}
\end{table*}

Moreover, 
the spin edge states accumulated by the SHE 
can be understood in the case of the Rashba spin-orbit
Hamiltonian,~\cite{sinova:126603}
by the deformation-induced gauge field, too.
The spin-orbit Hamiltonian in the SHE is given by
\begin{align}
 {\cal H}_{\rm so} = -\frac{\lambda}{\hbar}
 \bsigma \cdot (\hat{{\bf z}}\times {\bf p}),
 \label{eq:rashba}
\end{align}
where $\lambda$ is the Rashba coupling constant
and $\hat{{\bf z}}$ is the unit vector perpendicular to the plane.
First, we assume that 
the system is a cylindrical shape and periodic about $y$ direction.
Then we introduce the boundary at $y=0$ by replacing $p_x$ with 
$p_x - {\rm sign}(p_x) A^{\rm q}_x(y)$ 
in Eq.~(\ref{eq:rashba}) where $A^{\rm q}_x(y)(>0)$ is non-vanishing 
near the boundary $-\xi_g < y < \xi_g$
and ${\rm sign}(p_x)$ keeps the time-reversal symmetry.
The localized energy eigenstates can be obtained
as~\cite{sasaki06jpsj}
\begin{align}
 \psi_E({\bf r}) = 
 \begin{cases}
  \displaystyle 
  \exp\left( i \frac{p_x x}{\hbar} \right) e^{-y/\xi} 
  \begin{pmatrix}
   1 \cr 0
  \end{pmatrix} 
  & \displaystyle \left( y > 0 \right) \\
  \displaystyle 
  \exp\left( i \frac{p_x x}{\hbar} \right) e^{-y/\xi} 
  \begin{pmatrix}
   0 \cr i
  \end{pmatrix} 
  & \displaystyle \left( y < 0 \right),
 \end{cases}
\end{align}
where 
\begin{align}
 \frac{\hbar}{\xi}=p_x \tanh \left( \frac{{\rm sign}(p_x)}{\hbar} 
 \int_{-\xi_g}^{\xi_g} A^{\rm q}_x(y) dy \right).
 \label{eq:she_A}
\end{align}
Thus, by applying the electric field along $x$-direction, 
the initial extended state with $p_x=0$ ($\xi=\infty$) 
becomes $p_x\ne 0$ due to $dp_x/dt=-eE$, and 
can be localized.
This state can be considered as the critical state in the SHE.
The analogy between graphene and SHE systems is summarized in
Table~\ref{tab:analogy}.
It is interesting to see that
the Hamiltonian and time-evolution for polarization for graphene
and SHE have a special dual symmetry.

Albeit the similarity between the SHE and our system,
there are several differences.
First, by increasing the AB flux continuously 
to give a constant electronic field, 
the delocalization process occurs at the K' point.
It means that the pseudo-spin at the edge is not always increasing.
Second, the localization phenomena in our system 
depends on the shape of the edge, 
while such the structure dependent spin accumulation is not known 
for the SHE.
In our system, 
the dependence of the localization on the shape of the edge 
is given by ${\bf B}^{\rm q}({\bf r})$ field.~\cite{sasaki06jpsj}
To clarify this point, it is necessary to derive 
the deformation-induced gauge field for the SHE 
($A_x^{\rm q}(y)$ in Eq.~(\ref{eq:she_A}))
from a microscopic lattice model, 
which will be reported elsewhere.

The pseudo-spin accumulation may be useful 
like the applications for the SHE
since the presence of the edge states
is predicted to make the ferromagnetism 
in the presence of the Coulomb interaction.~\cite{fujita96}
Moreover, the electron-phonon interaction for 
the pseudo-spin polarized states is stronger than 
that for the extended states. 
The strong electron-phonon interaction may give rise to the 
superconducting states of the edge states.~\cite{sasaki06super}
Thus, we think that the coexistence of 
the localization transition described 
by the pseudo-spin accumulation
and real-spin polarization by Coulomb interaction 
will be an important subject of physics.

It is known that the next nearest-neighbor (nnn) hopping process 
gives a finite energy bandwidth for the edge states.~\cite{sasaki06apl}
Since the nnn hopping breaks the particle-hole symmetry,
the shift of the energy for the critical state
becomes either positive or negative value 
depending on the conduction or valence critical state, respectively.
Denoting the nnn hopping integral $\gamma_n$,
the shift of the critical state 
is given by adding $\pm c/L$ in Eq.~(\ref{eq:kc})
with $c \approx \gamma_n/\gamma_0$.
Theoretically, $c$ can be estimated around 0.1 
by Porezag {\it et al}.~\cite{porezag95}
Since $\gamma_n$ is renormalized
by the electron-phonon interaction, 
$c$ becomes much smaller than 0.1.~\cite{sasaki07local}
Thus, the change of $L$ in Eq.~(\ref{eq:L(d)}) due to $\gamma_n$ 
is less than 10 \% and is negligible.


In conclusion, 
we have shown that AB flux around 20[T]
induces localization-delocalization transition for the edge states
for metallic zigzag carbon nanotubes.
The localization is similar to the spin accumulation by the SHE
when we regard the pseudo-spin as the electron spin.
The LD transition can be observed by means of STM/STS.

\begin{acknowledgements}
 The authors would like to thank S. Murakami for valuable comments.
\end{acknowledgements}


\end{document}